\documentstyle[prl,aps]{revtex}

\input epsf
\begin{document}
\draft
% for two column  activate the line below...                
\twocolumn[\hsize\textwidth\columnwidth\hsize\csname@twocolumnfalse\endcsname

\preprint{} 
\title{Angular Position of Nodes in the Superconducting Gap of 
Quasi-2D Heavy-Fermion Superconductor CeCoIn$_{5}$}

\author{K.~Izawa$^{1}$, H.~Yamaguchi$^{1}$, Yuji~Matsuda$^{1}$, 
H.~Shishido$^{2}$, R.~Settai$^{2}$, and Y.~Onuki$^{2}$}
\address{$^1$Institute for Solid State Physics, University of Tokyo, 
Kashiwa, Chiba 277-8581, Japan}
\address{$^2$Graduate School of Science, Osaka University, Toyonaka, 
Osaka, 560-0043 Japan}
%\date{\today} 
\maketitle

\begin{abstract}

The thermal conductivity of the heavy-fermion superconductor 
CeCoIn$_{5}$ has been studied in a magnetic field rotating within the 
2D planes.  A clear fourfold symmetry of the thermal conductivity 
which is characteristic of a superconducting  gap with nodes along 
the ($\pm\pi,\pm\pi$)-directions is resolved.   The thermal 
conductivity measurement also reveals a first order transition at 
$H_{c2}$, indicating a Pauli limited superconducting state.  These 
results indicate that the symmetry  most likely belongs to 
$d_{x^2-y^2}$, implying that the anisotropic antiferromagnetic 
fluctuation is relevant to the superconductivity.  
    
\end{abstract}

% for two column  activate the line below... 
]

\narrowtext

%Introduction
The superconductivity with unconventional pairing symmetry has 
been a central subject of the physics of superconductors.  In the 
last two decades unconventional superconductivity has been found in 
several heavy fermion materials, such as CeCu$_2$Si$_2$, UPt$_3$, 
UPd$_2$Al$_3$, UBe$_{13}$ \cite{sigrist} and the recently discovered 
UGe$_2$\cite{saxena}.   There, the relationship between magnetism and 
superconductivity is the most important theme of research, because 
the strong electron-electron correlation effect which originates from 
the magnetic interaction between 4$f$ or 5$f$ moment and itinerant 
electrons allows a non-phonon mediated pairing with unconventional 
symmetry.  Very recently it has been reported that the family 
CeTIn$_5$ (T=Rh, Ir, and Co) are  heavy fermion superconductors 
\cite{petrovic,hegger}.  Especially, both CeIrIn$_5$ and CeCoIn$_5$ 
are ambient pressure superconductors, with transition temperature of 
0.4~K and 2.3~K, respectively.  Subsequent observations of power law 
temperature dependence of the specific heat \cite{petrovic,settai}, 
thermal conductivity\cite{movshovich}, and NMR relaxation 
rate\cite{kohori1,zheng} have identified CeTIn$_5$ as unconventional 
superconductors with line nodes.  The unique feature of these 
materials is that they bear some analogy with high $T_c$ cuprates.  
For example, the superconductivity appears in the neighbor of the 
antiferromagnetic state \cite{hegger}.  Moreover, the crystal 
structure is tetragonal which can be viewed as layers of CeIn$_3$ 
separated by layers of TIn$_2$ and electronic structure is quasi-2D 
\cite{petrovic,hegger,settai}.  Therefore they present an uncommon 
opportunity to study the unconventional superconductivity in the 
heavy fermion materials.
	 
Unconventional superconductivity is characterized by the 
superconducting gap structure which has nodes along certain 
directions.    Since the superconducting gap function is intimately 
related to the pairing interaction, its identification is crucial for 
understanding the pairing mechanism.  However, the detailed structure 
of the gap function, especially the direction of the nodes, is an 
unresolved issue in most of the  unconventional superconductors.   In 
fact,  to the best of our knowledge, it is only in high-$T_c$ 
cuprates and the B-phase of UPt$_3$ in which the nodal direction has 
been successfully specified.  The main reason for this is that the 
standard techniques to probe the unconventional superconductivity, 
such as penetration depth, specific heat,  ultrasonic attenuation, 
and NMR relaxation rate do not provide direct information on the node 
directions.   The most definitive test is a phase sensitive 
experiment which has been done for high-$T_c$ cuprates \cite{tsuei}.  
However, this technique appears to be available only for high-$T_c$ 
cuprates up to now.  The situation therefore strongly confronts us 
with the need for a powerful directional probe.

Recently it has been demonstrated both experimentally and 
theoretically that the thermal conductivity $\kappa$,  which responds 
to the unpaired quasiparticles (QPs) below $T_c$, is a powerful tool 
for probing the anisotropic gap structure 
\cite{yu,aubin,izawa,esquinazi,maki,vekhter1}.  An important 
advantage of the thermal conductivity is that it is indeed a {\it 
directional} probe, sensitive to the relative orientation among the 
thermal flow, the magnetic field, and nodal directions of the order 
parameter.  In fact, a clear 4-fold modulation of $\kappa$ with an 
in-plane magnetic field which reflects the angular position of nodes 
of $d_{x^2-y^2}$ symmetry has been observed in 
YBa$_2$Cu$_3$O$_{7-\delta}$ \cite{yu,aubin,esquinazi}, demonstrating 
that the thermal conductivity can be a relevant probe of the 
superconducting gap structure.  In this Letter, we have measured the 
thermal conductivity of CeCoIn$_{5}$ in magnetic field rotating 
within the 2D CeIn$_3$ planes.  Thanks to the two dimensionality,  a 
4-fold symmetry of $\kappa$ which clearly demonstrates the presence 
of gap nodes in the ($\pm\pi,\pm\pi$)-directions is observed.  We 
have also found the first order phase transition (FOPT) at $H_{c2}$ 
for the first time.  On the basis of these findings, we discuss the 
nature of the superconducting gap function of CeCoIn$_{5}$.
 
Single crystal CeCoIn$_{5}$ ($T_c$=2.3~K) was grown by the self-flux 
method \cite{settai}.  The residual resistivity ratio (RRR) was 
approximately 18.   The thermal conductivity was measured by the 
steady-stated method with one heater and two RuO$_{2}$ thermometers.  
The sample was cut into a rectangular shape 
($3.80\times0.38\times0.12$mm$^3$) and the heat current {\boldmath 
$q$} was applied along the [100] direction.  In the present 
measurements, it is very important to rotate {\boldmath $H$} within 
the 2D CeIn$_3$ planes with high accuracy because a slight 
field-misalignment produces a large effect on $\kappa$ due to the two 
dimensionality.  For this purpose, we used a system with two 
superconducting magnets generating {\boldmath $H$} in two mutually 
orthogonal directions and a $^{3}$He cryostat equipped on a 
mechanical rotating stage with a minimum step of 1/500 degree at the 
top of the Dewar \cite{izawa}.  Computer-controlling two magnets and 
rotating stage, we were able to rotate {\boldmath $H$} continuously 
within the CeIn$_3$  planes with a misalignment of less than 0.02 
degree from the plane, which we confirmed by the simultaneous 
measurement of the resistivity $\rho$.

We first discuss the $T$- and $H$- dependences of $\kappa$.    The 
inset of fig.~1(a) shows the $T$-dependence of $\kappa$ and $\rho$.   
Upon entering the superconducting state, $\kappa$ exhibits a sharp 
kink and rises to the maximum value at $T \sim$ 1.7~K.   The upturn 
of $\kappa$ is reminiscent of high-$T_{c}$ cuprates \cite{rcyu}. Similar 
$T$-dependence of $\kappa$ has been reported in 
Ref.\cite{movshovich}.   Compared to Ref.\cite{movshovich}, our 
$\kappa$ at the onset is slightly larger, but the enhancement below 
$T_{c}$ is smaller.   The Wiedemann-Franz ratio $L = 
\frac{\kappa}{T}\rho\simeq1.02L_0$ at $T_{c}$ is very close to the 
Lorenz number $L_{0}= 2.44\times 10^{-8}$~$\Omega$W/K, indicating 
that the electronic contribution well dominates over the phonon 
contribution.  Therefore,  the enhancement of $\kappa$ below $T_{c}$ 
is due to the suppression of the QP scattering rate,  similar to the 
high-$T_{c}$ cuprates. 

Figures 1(a) and (b) depict $H$-dependence of $\kappa$ for 
{\boldmath $H$}$\parallel ab$ ($H_{c2}\simeq 11$~T) and {\boldmath 
$H$}$ \perp ab$ ($H_{c2}\simeq 5$~T) below $T_c$, respectively.   At 
all temperatures,  $\kappa$ decreases with $H$ and the $H$-dependence 
becomes more gradual with decreasing 
\begin{figure}
\centerline{\epsfxsize 8cm \epsfbox{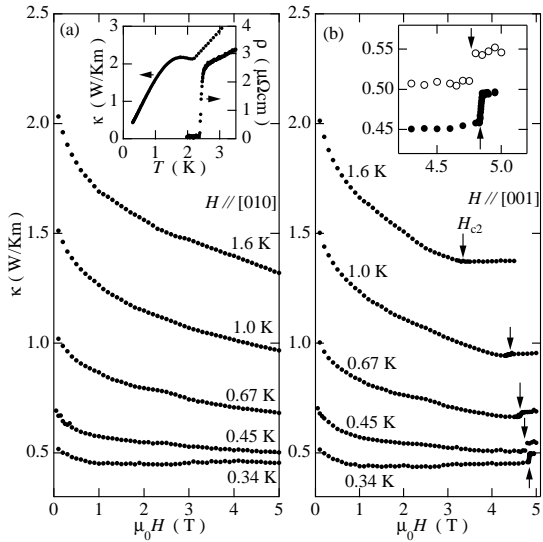}}
\caption{Thermal conductivity as a function of $H$ for (a) {\boldmath 
$H$}$\parallel [010]$ and (b) {\boldmath $H$}$\parallel [001]$ below 
$T_c$.  The thermal current {\boldmath $q$} is applied along 
[100]-direction.  Inset of (a) : $\kappa$ and $\rho$ in zero field.  
Inset of (b) : $H$-dependence of $\kappa$ near $H_{c2}$ at 0.45~K 
($\circ$) and 0.34~K ($\bullet$)}
\end{figure}
\noindent $T$ in both configurations.   
Interestingly, for {\boldmath $H$}$ \perp ab$,  $\kappa$ jumps to the 
normal state value at $H_{c2}$ below 1.0~K (see also the inset of 
Fig.~1(b)).    The magnitude of the jump in $\kappa$ increases with 
decreasing $T$.  Since the jump in $\kappa$  most likely comes from 
an entropy jump, this result provides a strong evidence of the 
occurrence of a first-order phase transition (FOPT).   As far as we 
know, {\it this is the first material which shows a FOPT at 
$H_{c2}$},  though a  FOPT is predicted to occur in the Pauli 
paramagnetically limited superconducting state \cite{fulde}.  We will 
discuss this subject later.
	
The understanding of the heat transport in the mixed state of 
superconductors with nodes has largely progressed during the past few 
years \cite{vekhter2}.   There, the dominant effect in a magnetic 
field is the Doppler shift of the delocalized QP energy spectrum, 
which occurs due to the presence of a superfluid flow around each 
vortex, and generates a nonzero QP density of states (DOS) at the 
Fermi surface \cite{volovik}.  While the Doppler shift increases 
$\kappa$ with $H$ through the enhancement of the DOS,  it can also lead 
to a decrease of $\kappa$ through the suppression of  impurity 
scattering time and Andreev scattering time off the vortices.  At 
high temperatures, the latter effect is predominant, while at low 
temperatures the former effect can exceed the latter effect,  as 
demonstrated in high-$T_c$ cuprates \cite{chiao}.  The data in 
Figs.~1 (a) and (b), in which the  $H$-dependence of $\kappa$ becomes 
more gradual with decreasing $T$, are consistent with the Doppler 
shift.  Thus at least above 0.35~K,  the Andreev scattering of the 
QPs is the main origin for the $H$-dependence of $\kappa$. 

We now move on to the angular variation of $\kappa$ upon the rotation 
of {\boldmath $H$} within the CeIn$_3$ planes.  Figure 2 displays  
\begin{figure}
\centerline{\epsfxsize 8cm \epsfbox{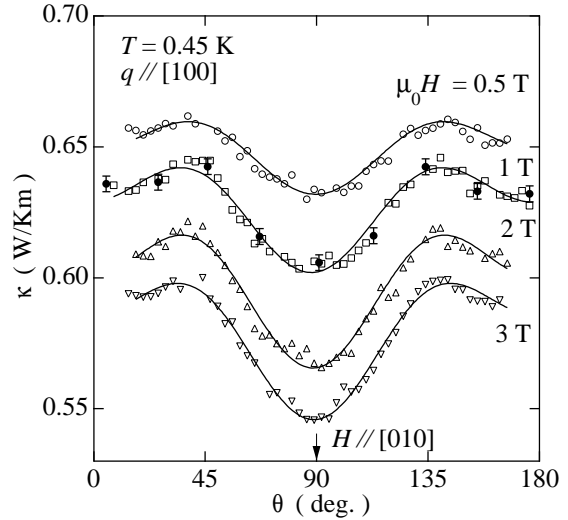}}
\caption{Angular variation ($\theta=$({\bf q},{\bf H})) of 
$\kappa(H,\theta)/\kappa_n$ at several $H$ for CeCoIn$_5$.  The solid 
lines represent the result of the fitting by the function 
$\kappa(H,\theta) = C_0 +C_{2\theta}\cos2\theta + 
C_{4\theta}\cos4\theta$, where $C_0$, $C_{2\theta}$ and $C_{4\theta}$ 
are constants.  The solid circles represent $\kappa(H, \theta)$ at 
$H$=1~T which are obtained under the field cooling condition at every 
angle.  For details, see the text.}
\end{figure}
\noindent 
$\kappa(H, \theta)$ as a function of $\theta=$({\boldmath $q$}, 
{\boldmath $H$}) at $T$=0.45~K, which is measured in rotating 
$\theta$ after field cooling at $\theta=0^{\circ}$.    The 
consecutive measurement inverting the rotating direction did not 
produce any hysteresis in $\kappa(H, \theta)$.  Moreover, the solid 
circles in Fig.2 shows $\kappa(H, \theta)$ at $H$=1~T which are 
obtained under the field cooling condition at every angle.   
$\kappa(H,\theta)$ obtained by different procedures of field cooling 
well coincide with each other.  Thus the field trapping effect 
related to the vortex pinning is negligibly small.   In all data, as 
shown by the solid lines in Fig.~2, $\kappa(H, \theta)$ can be 
decomposed into three terms with different symmetries;  
$\kappa(\theta) = \kappa_{0} + \kappa_{2\theta} + \kappa_{4\theta}$ 
where $\kappa_{0}$ is a $\theta$-independent term, and 
$\kappa_{2\theta} = C_{2\theta}\cos 2\theta$ and $\kappa_{4\theta} = 
C_{4\theta}\cos 4\theta$ are terms with 2- and 4-fold symmetry with 
respect to the in-plane rotation, respectively.   The term 
$\kappa_{2\theta}$, which has a minimum at {\boldmath 
$H$}$\perp${\boldmath $q$},  appears as a result of the difference of 
the effective DOS for QPs traveling parallel to the vortex and for 
those moving in the perpendicular direction.   
	
Figures 3 (a)-(d) display $\kappa_{4\theta}$ normalized by the 
normal state value $\kappa_n$  after the subtraction of the 
$\kappa_0$- and $\kappa_{2\theta}$-terms from the total 
$\kappa$.      It is apparent that $\kappa_{4\theta}$ exhibits a 
maximum at {\boldmath $H$}$\parallel$[110] and [1,-1,0] at all 
temperatures.    Figure 4  and the inset show the $T$- and $H$- 
dependences of $|C_{4\theta}|/\kappa_n$.   Below $T_c$ the amplitude 
of $\kappa_{4\theta}$ increases gradually and shows a steep increase 
below 1~K with decreasing $T$.  At low temperatures,  
$|C_{4\theta}|/\kappa_n$ becomes larger than 2\%.   It should be 
noted that this amplitude is more than 20 times larger than those of  
the 2D superconductor Sr$_2$RuO$_4$ with isotropic gap in the planes 
\cite{izawa}.  Then the most important subject is "Is the observed 
4-fold symmetry a consequence of the nodes?".   We here address the 
origin for the 4-fold symmetry.    There are several possible origins 
for this.  The first is the in-plane anisotropy of $H_{c2}$.   
According to Ref.\cite{settai}, $H_{c2}$ has very small but finite 
in-plane anisotropy; $H_{c2}\parallel [100]$ is approximately 2.7\% 
larger than $H_{c2}\parallel [110]$.   However, this anisotropy is 
too small to explain the large amplitude of $|C_{4\theta}|/\kappa_n>$ 
2\% at $H \ll H_{c2}$.     Further, and more importantly,  if this 
4-fold symmetry had come from the fact that $H_{c2}\parallel [100]$ 
is larger than $H_{c2}\parallel [110]$, the overall sign of this term 
should be opposite to the one actually observed in 
$\kappa_{4\theta}$.  The second possibility is the tetragonal band 
structure inherent to the CeCoIn$_5$ crystal.   If the in-plane 
anisotropy of the Fermi surface is large,  then the large anisotropy 
of $\kappa_{4\theta}$ should be observed even above $T_c$.  However, 
as shown in Fig.~4, the observed 4-fold symmetry above $T_c$ is 
extremely small; $|C_{4\theta}|/\kappa_n< $0.2~\%.  Thus the 
anisotropies arising from $H_{c2}$ and the band structure are 
incompatible with the data.  Moreover,  the amplitude of the 4-fold 
symmetry well below $T_c$ becomes more than 10 times larger than that 
above $T_c$.  These considerations lead us to conclude that {\it the 
4-fold symmetry with large amplitude well below $T_c$ originates from 
the QP structure.} 
	
We now address the sign of the 4-fold symmetry.  In the presence of 
nodes perpendicular to the layers, the term 
\begin{figure}
\centerline{\epsfxsize 8cm \epsfbox{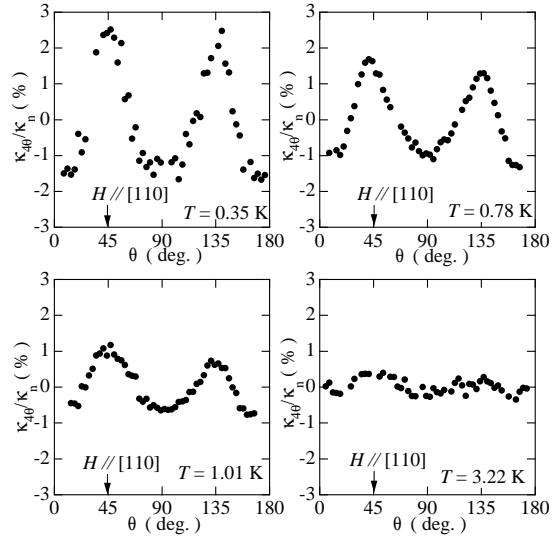}}
\caption{(a)-(d) The 4-fold symmetry $\kappa_{4\theta}/\kappa_n$ at 
several temperatures.  }
\end{figure}

\begin{figure}
\centerline{\epsfxsize 8cm \epsfbox{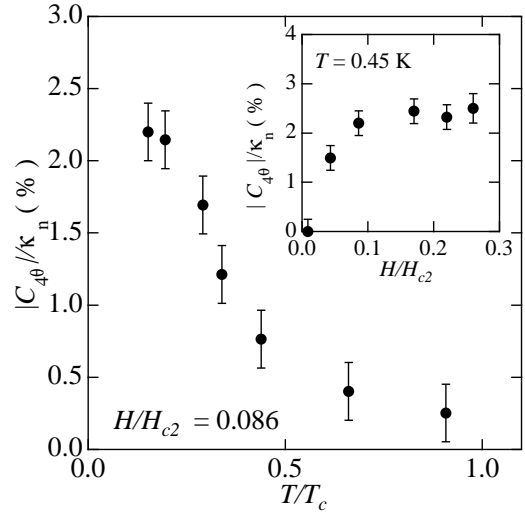}}
\caption{Amplitude of the 4-fold symmetry $|C_{4\theta}|/\kappa_n$ as 
a function of $T/T_c$.  Inset: Same data as a function of $H/H_{c2}$. 
}
\end{figure}
\noindent $\kappa_{4\theta}$ 
appears as a result of two effect.  The first is the DOS oscillation 
associated with the rotating {\boldmath $H$} within the 
$ab$-plane\cite{vekhter1}.  This effect arises because the DOS 
depends sensitively on the angle between {\boldmath $H$} and the 
direction of the nodes of the order parameter, because the QPs 
contribute to the DOS when their Doppler-shifted energies exceed the 
local energy gap.  The second effect is the the quasiparticle 
lifetime from the Andreev scattering off the vortex lattice, which 
has the same symmetry as the gap function \cite{yu,aubin,vekhter2}.  
As discussed before, the second effect is predominant in our 
temperature and field range.  In this case, $\kappa$ attains the 
maximum value when {\boldmath $H$} is directed to the nodal 
directions and becomes minimum when {\boldmath $H$} is directed along 
the antinodal directions \cite{yu,aubin,maki}.  Thus the sign of the 
present 4-fold symmetry indicates {\it the superconducting gap with 
nodes located along the ($\pm\pi,\pm\pi$)-directions, similar to 
high-$T_c$ cuprates}.
	
A quantitative comparison of the amplitude of the 4-fold symmetry 
with the theory reinforces this conclusion.  According to 
Ref.\cite{maki}, the 4-fold symmetry arising from the Andreev 
scattering off the vortices in $d$-wave superconductors is roughly 
estimated as 
$\kappa_{4\theta}/\kappa_n=-\frac{A(T)\sqrt{\pi}\Delta^2}
{\hbar^2\Gamma\varepsilon}\cos4\theta$.  
Here $\Delta$ is the superconducting gap, $\Gamma$ is the 
quasiparticle relaxation rate and 
$\varepsilon=\sqrt{2ev_fv'_fH_{c2}/\hbar}$ with $v_f$ and $v'_f$ the 
in- and out-of-plane Fermi velocity, respectively.  According to the 
numerical result, $A(T)$ is nearly zero at $T/T_c>$0.4 and shows 
rapid increase with decreasing $T$ at lower temperatures.  Similar 
tendency in the $T$-dependence of $|C_{4\theta}|/\kappa_n$ is 
observed, as shown in Fig.~4.  Using $\Gamma\simeq 
1.3\times10^{11}$s$^{-1}$, $v_f\simeq1\times10^4$m/s, 
$v_f\simeq5\times10^3$m/s\cite{settai}, $H_{c2}\simeq 11$~T, 
$2\Delta/k_BT_c$=3.54, and $A(T)$=0.033 at $T$=0.35~K from 
Ref.\cite{maki} gives $|C_{4\theta}|/\kappa_n\sim$8\%, which is in 
the same order to the data.  Thus Andreev scattering yields 
$|C_{4\theta}|/\kappa_n$ which is consistent with the data.  It is 
interesting to compare our results on CeCoIn$_5$ with the 
corresponding results on YBa$_2$Cu$_3$O$_{7-\delta}$, in which the 
4-fold symmetry has been reported in the regime where the Andreev 
scattering predominates.  The observed amplitude of 4-fold symmetry 
in YBa$_2$Cu$_3$O$_{7-\delta}$ is small; $\sim$ 0.4\% of total 
$\kappa$ at 6.8~K. However, this amplitude occupies a few \% in the 
electron thermal conductivity, because the phonon contribution is 
about 80-90\% of the total $\kappa$.  In YBa$_2$Cu$_3$O$_{7-\delta}$ 
with $d_{x^2-y^2}$ symmetry, $\kappa_{4\theta}$ has maxima  at 
{\boldmath $H$}$\parallel$[110] and [1,-1,0].  Thus 
$\kappa_{4\theta}$ of CeCoIn$_5$ is quantitatively in accord with 
$\kappa_{4\theta}$ of YBa$_2$Cu$_3$O$_{7-\delta}$.
	 
We finally discuss the symmetry of CeCoIn$_5$ inferred from the 
present results.  The fact that $H_{c2}$ is determined by the Pauli 
paramagnetic limit is a direct evidence of a {\it spin singlet 
pairing}, which is consistent with the recent Knight-shift 
measurements \cite{kohori2}.  Together with the fact that the 
superconducting gap has nodes at odd multiples of 45$^\circ$ in 
{\boldmath $k$}-space, we are naturally lead to conclude that {\it 
CeCoIn$_5$  most likely belongs to the $d_{x^2-y^2}$ symmetry} 
\cite{msr}.  The $d_{x^2-y^2}$ symmetry strongly suggests that the 
anisotropic antiferromagnetic fluctuation plays an important role for 
the occurrence of the superconductivity.  This  observation is in 
conformity with recent NMR and neutron scattering experiments which 
report  anisotropic spin fluctuation \cite{zheng,bao}.   
	
In summary, we have measured the thermal conductivity of the 
quasi-2D heavy fermion superconductor CeCoIn$_{5}$  as a function of 
the relative orientation of the crystal axis and the magnetic field 
rotating within the 2D planes.  A clear 4-fold symmetry of the 
thermal conductivity which is characteristic of a superconducting  
gap with nodes at odd multiples of 45$^\circ$ is revealed.  Rather 
surprisingly, we also observed a first order phase transition at 
$H_{c2}$ at low temperatures, indicating the Pauli paramagnetically 
limited superconducting state.  These results show that the symmetry 
of CeCoIn$_5$  most likely belongs to $d_{x^2-y^2}$, implying that 
the anisotropic antiferromagnetic fluctuation plays an important role 
for the superconductivity.   This material is the second example 
followed by high-$T_c$ cuprates, in which the nodal structure in the 
plane is successfully specified.  
	
{\it Note Added.} After completion of this work, we became aware of 
the work by T.~Sakakibara {\it et al.} who observed the FOPT at 
$H_{c2}$ by the magnetization measurements \cite{sakakibara}.  
	
We thank L.N.~Bulaevskii, T.~Kohara, Y.~Kohori, R.~Movshovich, 
T.~Sakakibara, M.~Sigrist, L.~Taillefer, A.~Tanaka, K.~Ueda, and 
I.~Vekhter for stimulating discussions.   We are also grateful to 
K.~Maki and H~Won for several comments and to T.~Sakakibara for 
showing us the unpublished data.

%%%


\begin{thebibliography}{9}
\bibitem{msr}       Here we disregard the the broken time reversal 
symmetry (BTRS).      If BTRS is present in CeCoIn$_5$, the gap 
function should contain an  additional imaginary term, such as 
$d_{x^2-y^2}$+$i\eta d_{xy}$ and  $d_{x^2-y^2}$+$i\eta s$ with very 
small $\eta\ll1$.   We note that  in CeIrIn$_5$ recent $\mu$SR 
measurement reported a presence of the BTRS \cite{heffner}, while 
another group reported an absence of BTRS \cite{higemoto}.
\end{thebibliography}
\end{document}